\documentstyle[multicol,pre,aps,epsf,amsmath]{revtex}

\begin{document}   

\draft
 
\title{Transitions to Line-Defect Turbulence in Complex Oscillatory Media}

\author{Andrei Goryachev$^1$, Hugues Chat\'e$^2$,  and Raymond Kapral$^1$}
\address{$^1$Chemical Physics Theory Group, Department of
Chemistry, University of Toronto, Toronto, ON M5S 3H6, Canada\\
$^2$CEA --- Service de Physique de l'Etat Condens\'e, 
Centre d'Etudes de Saclay, 91191 Gif-sur-Yvette, France}
\maketitle
\begin{abstract}
The transition from complex-periodic to chaotic behavior is investigated in 
oscillatory media supporting spiral waves. We find   
turbulent regimes characterized by the spontaneous nucleation, 
proliferation and erratic motion of synchronization defect lines
which separate domains of different oscillation phases. 
The line defect dynamics is controlled by the competition between
diffusion, which reduces line length and curvature, and 
phase-gradient-induced growth. The onset of each type of defect-line 
turbulence is identified with a non-equilibrium phase transition 
characterized by non-trivial critical exponents.
\end{abstract}


\begin{multicols}{2} 

\narrowtext 

Two-dimensional reactive media with oscillatory dynamics
support a variety of spatio-temporal patterns including spiral 
waves. In the vicinity of the Hopf bifurcation, spiral waves are
described by the complex Ginzburg-Landau equation (CGLE) \cite{kur,hagan}.
Spiral waves can also exist if the 
local dynamics is complex-periodic or even chaotic \cite{kle-BCM,PRE}. 
While the basic features of such regimes are 
akin to those of the CGLE,
a complete description of complex oscillatory media 
cannot be given in terms of the CGLE. 
For example, these media may undergo bifurcations 
where the period of the orbit doubles
at almost every point in space \cite{PRE,prl2}. The rotational
symmetry of spiral waves is then broken by the presence of 
synchronization defect lines where the phase of the local orbit
changes by multiples of $2\pi$. These defect lines have
been observed in a super-excitable system \cite{ijbc} and
in experiments on the Belousov-Zhabotinsky reaction \cite{jap}.

In this Letter, we study the fate of the synchronization defect lines as
the system parameters are tuned to approach the domain where spiral waves
have chaotic local dynamics. We show the existence of 
a new type of spatiotemporal chaos where the global temporal periodicity 
of the medium is broken by the spontaneous nucleation, 
proliferation and erratic motion of the defect lines separating
domains of different oscillation phases.
We describe the basic mechanisms governing the dynamics of the defect lines
and provide evidence that the onset of each type of defect-line turbulence
is a non-equilibrium phase transition with non-trivial critical exponents.
We also study inhomogeneous media without spirals where line motion has a 
different nature and different scaling laws due to the absence of overall 
phase gradients. 

We study reaction-diffusion (RD) systems where the local kinetics is described 
by  ${\bf R}({\bf c}({\bf r},t))$, a vector of nonlinear functions of the 
local concentrations ${\bf c}({\bf r},t)$. For simplicity, we
assume that all species have the same diffusion coefficient $D$. 
While our considerations should apply to any RD
system exhibiting a period doubling cascade to chaos, 
the calculations described here \cite{simdet} 
were carried out on the R\"{o}ssler model \cite{rossler}  
where $R_x = -c_y-c_z, \; R_y = c_x + Ac_y, \; R_z = c_x c_z - C c_z + B$. 
We investigate the behavior of the system as
$C$ increases, with other parameters fixed at $A=B=0.2$ and $D = 0.4$.

\begin{figure}[htbp]							       
\begin{center}
\leavevmode
\epsffile{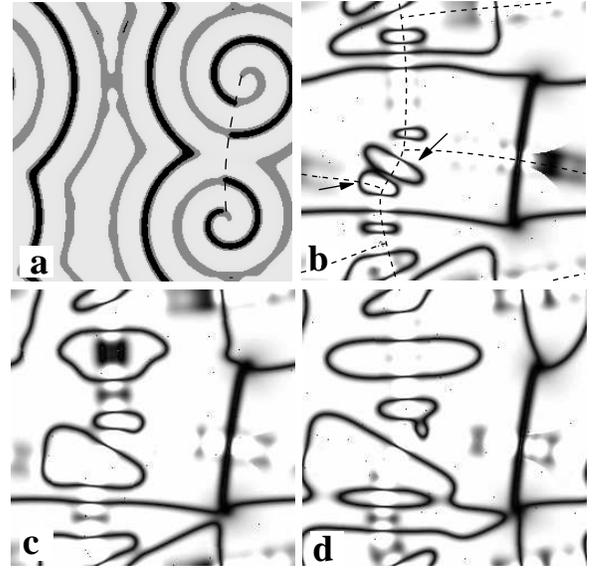}
\end{center}
\caption{The evolution of the medium in the $\Omega_2$-line turbulent 
regime at $C=4.38$. Square system of linear size $L=256$.
Panel (a) shows the $c_z({\bf r})$ field. The dashed line indicates
the $\Omega_1$ defect line. Panels (b), (c) and (d)
are equally separated in time by 48 spiral revolutions and show the 
$\xi_2({\bf r})$  field color-coded by grey shades. The fine dashed
lines in panel (b) mark the centers of the shock lines.}
\label{fig_6fr}
\end{figure}

Beyond the  Hopf bifurcation, the system supports a spiral solution 
and infinitely-many spatially-blocked configurations of spirals coexist
with spatially inhomogeneous states without spirals.
This multistability is preserved away from the Hopf bifurcation, 
even for $C$ values corresponding to chaotic regimes. 
The blocked configurations form irregular cellular structures, similar 
to those observed in the CGLE. 
Cells are centered on spiral cores and their polygonal boundaries 
are delimited by shock lines where spiral waves from two neighboring cells 
collide. Fig.~\ref{fig_6fr}(a) shows a snapshot of the 
$c_z({\bf r})$ field for a simple configuration with two spirals.

As $C$ increases beyond the Hopf point, the R\"ossler ODE system exhibits a period-doubling 
route to chaos followed by band-chaotic regimes intertwined with windows of periodic 
behavior. In spatially distributed media supporting spiral waves 
two period-doubling bifurcations take place at 
$C\simeq 3.03$ and $C\simeq 4.075$. These values are larger than the 
corresponding values for the ODE, 2.83 and 3.86, respectively. These
shifts in the bifurcation
diagram arise from the concentration gradients created by the spiral waves and
their values depend on the spiral wavelength. In spatially inhomogeneous media without
spiral waves the spatial gradients are small and  the shifts of the bifurcation points 
are not detectable. The period doublings in media with spiral 
waves are necessarily accompanied by the 
appearance of synchronization defect lines ($\Omega$ curves) whose 
existence is required to reconcile the doubling of the oscillation 
period and the period of rotation of the spiral wave \cite{PRE,prl2}.

In the period-2 regime ($3.03\le C \le 4.075$), 
a single type of synchronization 
defect line exists. These $\Omega$ curves are defined as the loci of those 
points in the medium where the two loops of the period-2 orbit
exchange their positions in local phase space and the 
dynamics is effectively period-1.
The period-2 oscillations on opposite sides of the $\Omega$ curve are shifted 
relative to each other by $2\pi$ (a half of the full period).
A medium with period-4 dynamics may support two types of synchronization 
defect lines corresponding to the two different possible types of loop 
exchange  for a period-4 orbit. Across $\Omega_1$ curves, 
which inherit properties of the $\Omega$ curves, 
the two period-2 bands of the period-4 orbits 
exchange, leading to a $\pm 2\pi$ phase shift across them.
The $\Omega_2$ curves correspond to the exchange of loops within the two 
bands, a finer rearrangement of the local cycle. 
Along them, the dynamics is effectively period-2 and there is a $4\pi$ phase 
shift as the curves are crossed.

Synchronization defect lines can be conveniently located by constructing
scalar fields encoding
the distance between loops of the period-4 orbit in local phase space.
To this aim, 
we chose to take advantage of the regular succession of peaks in the 
local time series of $c_z$, whose heights are in one-to-one 
correspondence with the various loops of the orbit. 
Calculating, at each point ${\bf r}$,
four such consecutive concentration maxima  $A_i({\bf r})$ and ordering them
so that 
$A_1({\bf r})\le A_2({\bf r}) \le A_3({\bf r}) \le A_4({\bf r})$,
one may construct the scalar fields  $\xi_1({\bf r})=A_4({\bf r})-A_1({\bf r})$ 
and $\xi_2({\bf r})=A_4({\bf r})-A_3({\bf r})$.
In the period-4 case,  $\xi_1({\bf r})$ and $\xi_2({\bf r})$ 
take on fixed non-zero values at points in the medium  
away from spiral cores and shock lines 
and vanish at points where the loop exchanges occur \cite{NOTE1}. 
Indeed, $\xi_1({\bf r})$ decreases to zero on the $\Omega_1$ curves
while $\xi_2({\bf r})$ vanishes on both the $\Omega_1$ and $\Omega_2$ curves. 
In the following, we study the fate of the $\Omega$ lines and 
use $\xi_1({\bf r})$ and $\xi_2({\bf r})$
both to determine their length and to visualize them. 
The $\xi_2({\bf r})$ field corresponding to Fig.~\ref{fig_6fr}(a) is shown in
Fig.~\ref{fig_6fr}(b). The thick vertical line connecting the spiral cores
is an $\Omega_1$ curve, while the thinner lines are $\Omega_2$ curves.

\begin{figure}[htbp]							       
\begin{center}
\leavevmode
\epsffile{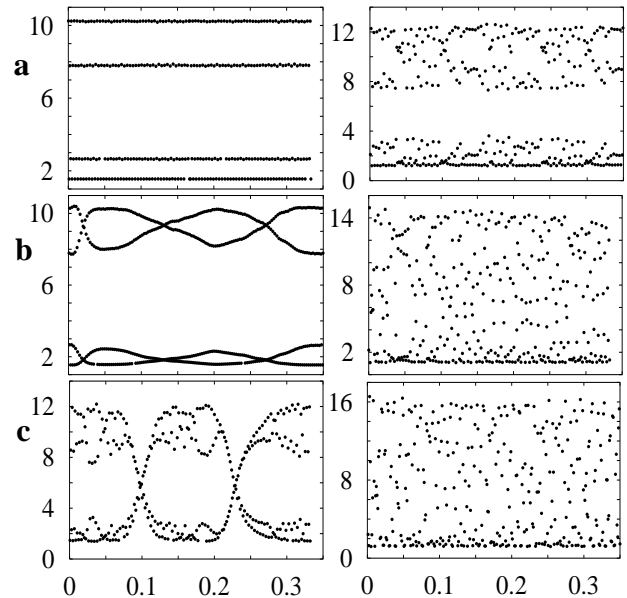}
\end{center}
\caption{Time series of $c_z$ concentration maxima: 
a) $C=4.30$, b) $C=4.42$ and c) $C=4.7$. Left panels show the local 
dynamics at a point in the cells, while the right panels show the dynamics 
on the shock lines. Time is in units of thousands of spiral revolutions.
In panels (b) and (c) the crossings of the $c_z$ maxima reflect 
the passages through the observation point of $\Omega_2$ or $\Omega_1$ curves, 
respectively.}
\label{fig_ld}
\end{figure}

On the shock lines, where the phase gradient vanishes, 
the local dynamics is approximately that of the R\"ossler ODE, and is thus
always more advanced along the bifurcation diagram. 
In particular, chaos first appears on the shock lines (for $C\simeq 4.20$). 
For $C=4.3$, where most of the medium is still in the period-4 regime,
two-banded chaos is seen on the shock lines (Fig.~\ref{fig_ld}(a)).
These localized chaotic regions give rise to fluctuations which 
may result in the creation of ``bubbles'' -- domains delineated by circular 
$\Omega_2$ curves (Fig.~\ref{fig_6fr}(b)).
For  $C \leq C_{\Omega_2}\simeq 4.306$, 
the bubbles are formed with a size smaller 
than a certain critical value and collapse shortly after their birth.
As $C$ increases beyond $C_{\Omega_2}$, 
the bubble nuclei begin to proliferate,  
forming large domains whose growth is limited by collisions with spiral 
cores or other domains. 

A typical life-cycle of a domain is illustrated in Fig.~\ref{fig_6fr}(b-d).
The shock lines are nucleation sites of $\Omega_2$ 
domains. Consider the two bubble-shaped nuclei 
indicated by arrows in panel (b) 
which were born in close proximity. In panel (c), they have coalesced, 
forming one rapidly-growing domain which then collides with its neighbor,
leaving a shrinking internal domain (panel (d)).
The contact of two $\Omega_2$ lines always leads to their reconnection and a  
reduction of their total length. The contact of $\Omega_1$ and $\Omega_2$ 
lines leaves one $\Omega_1$ line. This event is accompanied by
a change of sign of the phase shift across the $\Omega_1$ line.

\begin{figure}[htbp]							       
\begin{center}
\leavevmode
\epsfxsize=7truecm
\epsffile{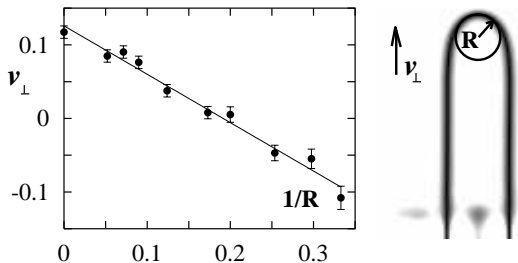}
\end{center}
\caption{Dependence of the domain propagation velocity, $v_{\perp}$,  
versus the curvature, $1/R$, of the tip for $C=4.30$.  A typical finger-shaped 
$\Omega_2$ domain is shown in the right panel along with the inscribed 
circle used to determine $R$.}
\label{fig_ve}
\end{figure}

The evolution of the size and shape of the $\Omega_2$ line
encircling a domain is controlled by the balance of two 
competing factors: propagation along phase gradient 
directed toward spiral cores 
which results in line growth, and the tendency of 
diffusion to eliminate curvature and reduce the length of defect lines. 
To investigate the interplay of these two factors, 
a series of simulations was carried out
on a system  without spiral waves, but with  constant 
concentrations corresponding to those in the spiral core  
imposed along one pair of parallel boundaries.
This effectively creates ``spiral core'' boundaries emitting
trains of plane waves which
collide in the center of the system to form a straight shock line. 
In this case, the nucleated domains
have very simple finger-like shapes, normal to the shock line, and consist 
of two straight segments 
with approximately semi-circular caps (Fig.~\ref{fig_ve}).
The domain growth velocity normal to the core boundaries,  $v_{\perp}$,
varies with the radius $R$ of the arc of the growing tip as
$v_{\perp} = v_p - \Delta/R$
where $v_p\simeq 0.126$ is the velocity of a straight $\Omega_2$ line 
parallel to the core boundaries, and $\Delta\simeq 0.658$. 
The linear dependence of $v_{\perp}$ on $1/R$ shown in Fig.~\ref{fig_ve} 
confirms the effect of mean curvature on the velocity of $\Omega$ line
propagation. Since the width of small domains is approximately 
equal to $2R$, one can estimate
the critical size that must be exceeded for 
domain proliferation. One finds $R_c\simeq 5.228$,
in good agreement with direct measurements from the 
observation of domains whose shape does not change with time.

The transition to $\Omega_2$-line-defect turbulence in media with
spiral waves, which occurs around
$C=4.3$, changes the character of the local 
dynamics observed in the bulk of the medium. As the $\Omega_2$ lines 
propagate, the associated loop exchanges result in an 
effective band-merging in the orbits of local trajectories so that they take 
the form of two-banded chaotic trajectories (Fig.~\ref{fig_ld}(b)). 
Although the local trajectories retain their period-4 structure 
between two passages of $\Omega_2$ lines,
the long-time trajectory cannot be distinguished 
from that of two-banded chaos. Thus, the global transition of the medium 
to defect-line turbulence can be characterized locally as intermittent 
band-merging. 

As the parameter increases further ($C>4.44$) the local 
dynamics undergoes prominent changes. It fails to 
exhibit a period-4 pattern in the intervals separating line defect passages, 
and consists instead of four-banded 
orbits whose bands grow in width 
with increasing $C$ and merge at $C\gtrsim 4.7$.
Together with this permanent band-merging,
spontaneous nucleation of $\Omega_2$ bubbles occurs in the bulk.
These chaotic $\Omega_2$ lines are the loci of medium points 
where the two chaotic bands of the 
local orbit shrink and a thick ``period-2'' orbit is formed. 
As $C$ increases, their width decreases,  
and for $C\gtrsim 4.8$ the $\Omega_2$ lines cease to exist as well
defined objects.

While the local dynamics changes continuously
to four- and subsequently two-banded chaos, 
another transition, mediated by moving $\Omega_1$ lines, takes place. 
At $C=4.557$, the shock regions, where the local dynamics 
exhibits one-banded chaos, begin to spontaneously nucleate 
bubbles delineated by $\Omega_1$ lines. As $C$ increases, the newly-born
domains begin to proliferate.
The qualitative features of this transition are similar 
to those of the $\Omega_2$-line turbulence transition:
the dynamics of the $\Omega_1$ lines encircling domains 
is controlled by the factors discussed above and, considering shape of the
long-time local phase space trajectories, it can be associated with 
intermittent band-merging (Fig.~\ref{fig_ld}(c)),
leading to one-banded local chaotic orbits.

As the parameter $C$ increases even further, beyond $C \geq 5.0$, the
local trajectories in the bulk of the medium 
exhibit complete band-merging to one-banded chaos. In this
regime,  no defect lines can be identified and
further increase in $C$ does not result in any qualitative changes. 
However, spiral waves continue to exist, 
signalling the robustness of phase synchronization 
in this amplitude-turbulent regime. \cite{pik}

\begin{figure}							       
\begin{center}
\leavevmode
\epsfxsize=8.5truecm
\epsffile{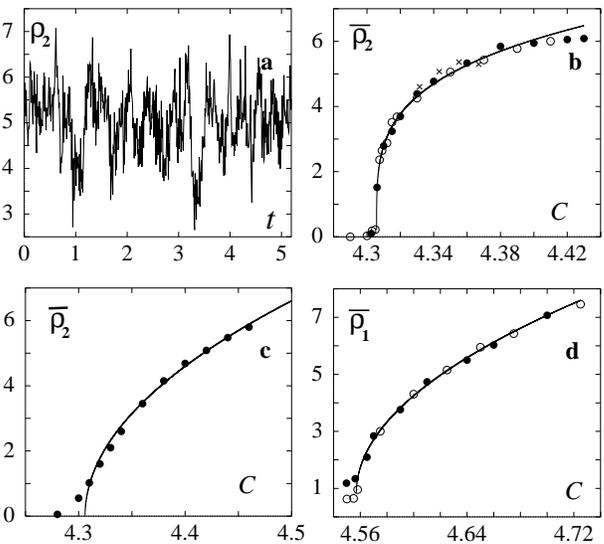}
\end{center}
\caption{Critical properties of the two defect-line mediated phase 
transitions observed. (a) Time series of $\rho_2$ at $C=4.34$
for the configuration shown in Fig.~\protect\ref{fig_6fr}
($L=256$, two spirals; time is in units of thousands of spiral revolutions). 
(b) Variation of $\bar{\rho_2}$ with $C$ ($\circ$ -- $L=256$, 
two spirals; $\bullet$ -- $L=256$, four spirals; 
$\times$ -- $L=512$, two spirals). 
(c) $\bar{\rho_2}$ versus $C$ for an inhomogeneous system without 
spiral waves, $L=256$. 
(d) Same as (b) but for the $\Omega_1$-line transition.
The solid lines are power-law fits.
}
\label{fig_tr}
\end{figure}

We now focus on the two onsets of synchronization defect line turbulence.
The $\Omega_i$ line density, $\rho_i(t)=\ell_i(t)/\sqrt{S}$, where $S$ 
is the surface area of the medium and  $\ell_i(t)$ the instantaneous total length
of $\Omega_i$ lines, can serve as an order parameter to characterize
these transitions. 
Above each transition threshold, and as long as the 
corresponding defect lines continue to exist,
the balance between line growth and destruction
results in a statistically stationary average density $\overline{\rho_i}$, 
while  $\rho_i(t)$ fluctuates. Thus the time series of $\rho_2(t)$ above the first threshold 
shown in Fig.~\ref{fig_tr}(a) 
exhibits high-frequency, low-amplitude fluctuations attributed to
the birth and death of nuclei in the shock regions, as well as large-amplitude 
oscillations with long correlation time. This suggests that the proliferation 
of domains and their destruction through coalescence occurs cooperatively.
This is confirmed by the fact that,
for both transitions, the order parameter goes continuously to zero
as $C$ decreases toward the threshold. (In Fig.~\ref{fig_tr}(d) the 
$\bar{\rho_1}$ density does not vanish below threshold because the
contribution from the stationary $\Omega_1$ line shown in 
Fig.~\ref{fig_6fr} has not been removed.) The data fall on curves with 
power-law forms, $\overline{\rho_i}(C) \sim (C-C_{\Omega_i})^{\beta_i}$,
the signature of continuous phase transitions.
The critical values are found to be
$C_{\Omega_2} \simeq 4.306$ and $C_{\Omega_1} \simeq 4.557$,
while the critical exponents are
$\beta_2 \simeq 0.25$ and $\beta_1 \simeq 0.49$.
Finite-size effects usually accompany
critical point phenomena as correlation lengths diverge near threshold.
Here, the finite-size to consider is the typical
size of the cells composing the spiral wave structure.
Strictly speaking, $\bar{\rho_i}$ is not an intensive quantity because
line-defect motion is constrained to occur between 
the network of shocks (where they nucleate) and the spiral cores, and
this area varies from one spiral configuration to another.
However, the data in Fig.~\ref{fig_tr}(b,d) show that $\bar{\rho_i}(C)$ 
depends weakly on the cell size. 

A number of conclusions can be drawn from the simulation results. 
The two transitions exhibit significantly different scaling 
properties, a remarkable fact given that the mechanisms at play appear
to be the same in both cases. This difference may arise from the fact that 
the $\Omega_1$ line transition takes places
when $\Omega_2$ lines still exist in the medium. 

The zero-spiral-density limit is singular since the transitions
observed in the medium  without spiral waves are different from those
described above. In this case, the onset of 
defect-line nucleation occurs at the same critical values $C_{\Omega_i}$.
However, in the absence of large-scale phase gradients,	   
the entire medium behaves like the shock regions separating spiral 
wave cells, defect lines do not grow and the increase of $\overline{\rho_i}(C)$ 
arises essentially from the enhanced nucleation rate. 
This leads to a different form of the onset of line turbulence 
(cf. Fig.~\ref{fig_tr}(c) for the behavior of $\bar{\rho_2}(C)$) 
characterized by different critical exponents ($\beta^*_1=1.22$,
$\beta^*_2=0.53$). These values are difficult to estimate because of 
fluctuations in the 
$\xi_i$ fields not associated with fully developed $\Omega$ lines.  
The fact that $\beta_1^*$ and $\beta_2^*$ are significantly different from 
$\beta_1$ and $\beta_2$ supports the conclusion that the character of the 
transitions is different in the spiral and spiral-free systems.

To our knowledge, 
there is no equilibrium equivalent of these phase transitions, nor were 
their non-equilibrium analogs reported previously. 
The line-defect phase transitions may constitute a special class of 
non-equilibrium critical phenomena since, in this form of spatiotemporal 
chaos, it is the dynamics of one-dimensional synchronization defects that 
breaks the global temporal periodicity of the medium. 
	     
Finally, 
the observations of synchronization defect lines in a number of excitable 
systems \cite{ijbc,jap} demonstrate their independence
of any particular reaction mechanism. Therefore, the phenomena
presented in this Letter should be observable not only 
in chemical media exhibiting period-doubling
but also in a much broader class of systems. 
For example, they may exist in the cardiac muscle
where complex-excitable dynamics and spiral waves, necessary prerequisites
for the emergence of synchronization defects, 
have been established experimentally \cite{card}.

\end{multicols}


\begin{thebibliography}{99}
\bibitem{simdet} An explicit Euler scheme with spatial derivatives given 
by simple second-order finite differences was used to integrate the PDE.
The time step $\Delta t = 10^{-2}$ (about 1/600 of the basic oscillation
period), space step $\Delta r = 1.$, and the scaled diffusion coefficient 
$D \Delta t/ (\Delta r)^2 = 4\times 10^{-3}$ were the same for all 
calculations.
\bibitem{NOTE1} Determining defect lines via the $\xi$ fields 
is possible because
the dynamics of these curves is much slower than the overall oscillation
of the medium.
\end{thebibliography}
\end{document}